\documentclass[singlespacing]{elsart}
\usepackage{epsfig}
\usepackage{amssymb}

\begin{document}

\begin{frontmatter}

\title{Departure From Gaussian: The Waiting Time Statistics of X-Ray Solar Flares}
\author{Deborah Leddon}

\address{University of North Texas, Center for Nonlinear Science,
Department of Physics, Denton, Texas  76203}

\begin{abstract}
Previous studies of the statistical behavior of solar flare waiting
times are based on the assumptions of Gaussian
and Poisson statistics,
subject to central limit theorem restrictions.
In this study, the results of a rescaled range analysis on the waiting times 
for two hard x-ray solar flare data sets are presented. The Hurst 
scaling parameter, H,  for both data sets is well above 0.5, clearly indicating that the
statistics of the data has departed from ordinary Brownian motion and is 
characterized by memory correlations. In addition, the distribution
exponents, $ \mu $, when compared to the same values obtained from the relationship between the scaling parameter and $ \mu $
for an asymmetric jump model, reveals
that the waiting time distributions are characterized by L\'{e}vy statistics. 

\end{abstract}

\begin{keyword}
x ray solar flares, FBM, L\'{e}vy distribution, Scaling, R/S analysis.
\PACS 05.40.+j; 95.10.F; 96.60Rd 
\end{keyword}
\end{frontmatter}

\section{Introduction}

Solar flares are transient and highly explosive energy events which emit radiation spanning
the entire electromagnetic
spectrum as well as high speed particles.
Large ranges of time
and spatial scales as well as energies up to $10^{32}$ ergs are involved. Since
flares generally occur
within the more active regions of the sun which contain magnetic bipolar
areas, the generation and release of large amounts of energies and
particles is inferred to
be magnetic in origin. Among the many dynamical models around for
explaining the flaring process the
most generally accepted one is the 'Standard Model', (page 282; Golub) \cite{golub} which
depicts flares as sudden releases of
superheated particles and plasma radiation generated by the
reconnection of the bipolar field
upon recovery from the non-equilibrium state. These events
are often hard to predict due to the  
complexity and chaotic nature of the interactions between the solar magnetic field with the 
surrounding coronal medium, phenomenon which are
not well understood at present.
To complicate the picture even further, it is possible that
the dynamics of the flaring process may
not be fully observable if the initial triggering
instability itself lies buried in background noise.

It is reasonable then that investigators turn to statistical
methods in hopes of making sense of flaring behavior. To date,
these methods have generally been based on the assumption that
fluctuations
leading to flare generation and energy release are independent,
random processes that can be
characterized by the finite moments of distributions subject to
the restrictions of the Central Limit
Theorem \cite{reich}. Two of these distributions, the Gaussian and it's related distribution, the Poisson,
are most often used in the analysis
of flare time series. A major implication of this use, however, is
that the frequency distribution exponents and hence the scaling of the time series
being analyzed is often measured incorrectly or is obscured. 

It is now generally accepted that most
physical phenomena do not evolve according to well defined inventories
of physical processes but rather exhibit characteristics of systems
sensitive not only to the interactions of internal fluctuations but
to environmental ones as well. The composite traits of randomness
and order displayed by these systems
are evident in the invariant scaling of the system's statistical
behavior in both the temporal and spatial scales and by moments
which are either non-finite or do not exist at all. In addition,
the ubiquity of the inverse power law, which is noted not only
in flare observations but elsewhere in the observation and
analysis of other physical phenomena is an example of this
universal behavior. Examples of other nonlinear systems which
scale in such a way include stockmarket indices \cite{mantegna}, heart
interbeat intervals\cite{west}, teen birth behavior \cite{west2}, DNA sequences \cite{allegro},
and more. 

Current efforts to relate frequency distributions to the
underlying dynamics of the flaring process have involved not
only the assumptions of gaussian and/or Poisson statistics but also the fitting
of these distributions to data. In the
case of waiting time distributions, results differ. For example,
the waiting time distribution of hard x-ray flare events recorded
for 8 years by the ICE/ISEE3 spacecraft and examined by Wheatland
et al \cite{wheat} as a Poisson process, evidenced correlations between events,
particularly an overclustering with respect to short waiting times.
Normally this is taken as proof of sympathetic flaring; i.e.; where
one flaring event triggers another, however, the authors were
unable to distinguish between this result and the possiblity that
individual flares might be comprised of several bursts events. Biesecker \cite{doug},
using using one year of x-ray data from the BATSE catalog found
a waiting time distribution based on a time varying poisson in
which the mean flaring rate was not finite. More recently, waiting
time distributions extracted from 20 years of GOES x-ray flaring
data were examined by Boffetta et al \cite{guido} who determined that the inverse 
power law signature was indicative of MHD turbulence and in contradiction with the time dependent
Poisson statistics of Wheatland \cite{wheat2} for the same set of data. 
Aschwanden et al \cite{asch}, went one step further by implementing
the logistic function,
as the form of the dissipated energy rate, an assumption,
  which does in part address the obvious stochastic nature
  through the evolution of cascading bifurcations
  The fluctuations were 
again Poisson distributed with frequency distributions seen as 
inverse power law or exponential in form. The interpretation 
of results such as inverse power law forms has often been seen
as evidence that the underlying dynamics is manifested as a self organized state
which has exceeded it's critical limits of stability. 

It is desirable then to take note of some of these results;
the time varying Poisson, the non-finite means and the inverse
power law signature, all of which clearly indicate the need to
extend the basic assumptions of random gaussian statistics in
the analysis of solar flare time series.

The waiting time statistics are derived from times corresponding to the
peak intensities of hard x-ray flares/bursts events which occur in the impulsive phase of a total flaring event. 
The data is comprised of 
nonlocalized flares representing the entire flaring behavior of the Sun. Hence, 
only part of the dynamical picture of single flaring events are being analyzed within the overall context 
of an aggregated flaring system. We examine the waiting time distributions of
two data sets of X-ray solar flares in order to make clear the departure of these flares from the random gaussian regime
as well as examining the scaling characteristics. 

The Rescaled range analysis ($ R/S$) technique was applied to the waiting time data sets in 
order to determine the scaling parameter, H, known as the 'Hurst exponent' and examine 
the degree of departure from the random gaussian regime. This examination assumes Fractional Brownian 
Motion (FBM) statistics in which the individual fluctuations (data points) are independently
gaussian distributed. This paradigm is a generalization of ordinary brownian motion in which
Mandelbrot \cite{feder} incorporated the random constant, $ H =  0.5$, of the second moment into a range of values
accounting for persistant and anti-persistant biased behavior noted in the $ R/S$
analysis of a number of phenomenon. While the $ R/S$ analysis of an FBM process is extremely robust, even
for non-gaussian independent processes described by a log-normal, hyperbolic, or gamma distributions, \cite{feder}, it does not
make a distinction between fractional gaussian and fractional non-gaussian statistics. Hence, despite the theoretical assumption
of gaussian statistics representing the dynamics of an observed process, it may still be the case that the underlying distributions are non-gaussian.

The rest of the paper is outlined as follows: section two describes
the rescaled range theory under the assumptions of the FBM paradigm, section three discusses the analysis and results, 
and section four summerizes the conclusions.
\section{Theory: Rescaled Range Analysis and Fractional Brownian Motion}
The assumption of gaussian fluctuations in nonlinear analysis techniques can be a useful one especially if
the moments of the data are the analyzed parameters. In our analysis of x-ray flare waiting times, it is sufficient 
to examine the scaling behavior of the accumulated departures of the mean fluctuations (waiting time and peak intensities data points) within 
windowed partitions of the data which are then normalized by the standard deviation. Any
long term memory effects are evidenced in these accumulated mean departures. For a data set containing the fluctuations; $ \xi _{i}$, 
we denote the mean value
\begin{equation}
< \xi >_{\tau } = \frac{1}{\tau } \sum_{i= 1}^{\tau } \xi _{i\left(t \right)}
\label{eq1}
\end{equation}
in order to define a collection of departures from the mean value for windows of the data, $ t$,
\begin{equation}
X\left(t,\tau  \right)= \sum_{i= 1}^{t} \left(\xi _{i}- < \xi >_{\tau }  \right).
\label{eq2}
\end{equation}
Next, the range of differences, $ R\left(\tau  \right)$ between the maximum and minimum values of X(t) is defined as 
\begin{equation}
R\left(\tau  \right)= \max_{1\leq t\leq \tau}  X\left(t,\tau  \right)  - \min_{1\leq t\leq \tau} X\left(t,\tau  \right).
\end{equation}
Finally, we divide the $ R\left(\tau  \right)$ by the standard deviation, $ S\left(t,\tau  \right)$, for each partition
\begin{equation}
S\left(t,\tau  \right)= \left[\frac{1}{\tau }\sum_{i= 1}^{\tau } \left(\xi _{i}- < \xi > _{\tau } \right)^{2}  \right]^{\left(1/2 \right)}.
\label{std}
\end{equation}
Note, that since different values of $ R\left(t,\tau  \right)$ for each interval start at different times, t, 
the division by the standard deviation normalizes and rescales the range of fluctuations within the same time period.
The rescaled range relationship is then defined as 
\begin{equation} 
R\left( \tau \right)/S\left(\tau  \right)= \left(a \tau  \right)^{H}
\label{rovs}
\end{equation}
H is known as the Hurst exponent after H. E. Hurst, an  hydrologist for the Aswan dam project, who originally developed the 
method in order to determine ranges of Nile river flow rates for the purposes of designing a resevoir. In 
using over a thousand years of river flow level data, he discovered a non-random pattern of high flow and low flow levels \cite{hurst}. 
A number of studies completed on other rivers revealed a range of decimal values above 0.5 but always lower than 1.0. He concluded that 
some type of persistant memory effect was at 
work in the relationship of the river dynamics with the outside environment. Mandelbrot and Wallis \cite{mandl} extended the 
rescaled range technique by incorporating the Hurst 
exponent as a scaling parameter. Retaining gaussian increments 
but with an inverse power law signature to account for scaling, the H parameter can reveal three different regimes of scaling. 
For the 'persistant' regime, $ 0.5 < H \leq 1.0$, 
data fluctuations will be positively correlated, meaning that a fluctuation increment 
in one direction  is likely to be followed by another fluctuation increment in the same direction. For the 'anti-persistant'  
regime, $0 <  H < 0.5$, the opposite effect is achieved; fluctuation increments in one direction are likely to be followed by increments 
in the opposite direction. Regular Brownian motion is recovered for H = 0.5, a random walk situation in which no memory effects 
can be present.

Use of the rescaled range analysis technique to uncover long term memory signatures in solar activity
began with the analysis of average monthly sunspot numbers yielding H values of 0.9 indicating a high level of 
memory \cite{mandelbrot}. Other studies, based on Carbon 14 solar proxy data \cite{ruz} and solar doppler rotation data \cite{komm} yielded Hurst values indicating
a high degree of persistance. Recently, a Hurst value of 0.74 was determined for the Hydrogen alpha flaring index, Q,
a measure of flaring activity which correlated well with the measure of sunspot activity \cite{lepr}.
\section {Analysis and results}
We examine the waiting time behavior of two sets of solar flare data. The ICE/ISEE3 (International Cometary Explorer) spacecraft data 
comprises x-ray data containing the times of x-ray peak flux events taken from August, 1978 to May, 1986.  This data set, comprised of 3574 events and 
used by a number of solar researchers \cite{bromund}, is an eight year series of uninterrupted data. The second set of data 
comes from the solar flare x-ray catalog list obtained from the Burst and Transient Source experiment (BATSE) 
onboard the Compton Gamma Ray observatory (CGRO) satellite and represents the corresponding times of 
peak flux counts from x-ray events listed from 1991 to 2000. The total number of data points is 7212. The waiting times are defined as the laminar lengths
between two nearest neighbor times corresponding to successive peak bursting events. Both data sets
represent the entire sun system in that they consist of data taken from non-localized flaring events.

A rescaled range analysis 
was performed on the waiting times through non-overlapping windows of the data points. Based on equation (\ref{rovs}), 
a linear regression of the log $ R/S$ values versus the log of the windowed times, $ \tau $, was then performed to determine 
the H values from the slopes. Figures 1 and 2 show the final results. It should be noted that the regression fit of the ICE/ISEE3 $ R/S$ values was performed through 
the linear region (first 20 points) of the data which resulted in a higher H value of
$ 0.72\pm 0.01$ for the waiting times. This restricted fit was due to the unstable fluctuations of the data after the break in the curve at around the 
twenty first point. This can be seen in figure 1, where quite clearly, a fit through the unstable regions would have 
given incorrect H values. The regression fit in figure 2 involved approximately, the same first 20 points. In each plot, a solid line representing a randomized set of 
data for which $ H= 0.5$ (scaled to the minimum data window value), is included to illustrate the departure of the intensities and 
waiting times from the random gaussian regime. 

How can we be sure if the waiting times parameters of our flaring data correspond to 
dynamical processes with a long time memory? The answer to this question lies
in the assurance that our calculated H values adequately reflect the memory signature of the underlying dynamics.
In order to assure the statistical significance of the H values we have calculated, we employ the use of the 
surrogate (randomized) data technique and hypothesis testing. We obtain eleven sets each of randomized waiting times from both flare 
data sets, by employing the use of an iterated random seed function written in  
Mathematica 4.0 code. Then, the rescaled range analysis of each randomized data set was performed in order to obtain
eleven H values for each set of waiting times and peak intensities. We calculate the means of the four sets of H values and
define a significance level, $ \alpha = 0.01$, (confidence interval = $ 1- 100\alpha $) against which 
the probabilities (denoted as "p values") of these means are tested. Our hypothesis test is a standard two sided one \cite{harnett} in which we
test whether the random H values can have a mean value other than 0.5. Said more simply, if we obtain probability 
values that are smaller than the p value defined by $ \alpha $, then we reject the null hypothesis that 
the means of the randomized H values can be one of the H values greater than 0.5. This can easily be seen in Table 1  
which summarizes the H values of the non-randomized data and their corresponding p values.
Since all of the p values are extremely small based on the 0.01 significance level, we know
our H values to be statistically significant. Hence, we have correctly determined that the underlying dynamics
responsible for the temporal evolution of hard x-ray flaring processes have long time memory correlations
and are not random gaussian processes.

At this point we focus on the relationship between the frequency distribution exponents and the H values in order to 
examine the statistics of the underlying flaring process. We denote the waiting times of the ICE/ISEE3 and BATSE flares 
as $ \psi_{I} \left(t \right)$ and $ \psi_{B} \left(t \right)$ respectively. Utilizing the assumptions of the CTRW under the 
fractional brownian motion paradigm, we employ a more generalized form of the inverse power law;
\begin{equation}
\phi \left(y \right)= \frac{A}{\left(B+ y \right)^{\mu }}
\label{invpwr}
\end{equation}
where the normalization condition,
\begin{equation}
A= \left(\mu - 1 \right)B^{\mu - 1}
\end{equation}
implies two y parameter regions; the small parameter region, $ y\ll B$, and the stationary
asymptotic region, $ y\gg B$. Note that the most utilized form of the inverse power law,
\begin{equation}
\phi \left(y \right)= \frac{A}{y^{\mu }}
\end{equation}
is recovered for the asymptotic assumption. Previous studies on waiting time distributions \cite{bettin} 
have utilized the generalized form for 
$ y= t$ in order to make use of its integrable properties in the continuous time random walk scheme.
The motivation for examing our solar frequency distributions in terms of this form is two fold;
it is a more realistic form exhibiting the dual $ \psi \left(t \right)$  parameter 
region appearance of these frequency distributions and to stay within the CTRW scheme.

Figures 3 and 4 depict the frequency distributions and the fits based on the generalized inverse power law.
The use of the generalized inverse power law form allows for the recovery of distribution exponents which are 
larger than their asymptotic inverse power law equivalents. The larger values are due to the 
contribution of the smaller waiting times parameter regions. By comparison, using the same ICE/ISEE3 data, 
Wheatland et. al. \cite{wheat} that 
for $ \psi_{I} \left(t \right)$, $\mu =1.4$. For peak intensities data obtained from the BATSE database
for the period, 1991-1992, Biesecker found $\mu = 1.68\pm 0.02$ \cite{doug}. 

Remembering that our FBM picture is not complete in that the persistant region is characterized only
by gaussian fluctuations, we utilize the H scaling parameter as 
an approximation to the true scaling when attempting to characterize the underlying statistics. Studies within the last decade
indicate that this region of persistance is characterized by processes which which diffuse faster
than what an ordinary random walk model can replicate. What is known is that systems
characterized by scaling behavior may exhibit intermittent and bursting type behaviors which can
only be modeled as processes diffusing faster than ordinary. From a random walk point of view, what 
was once a walk is no longer sufficient to explain jumping or bursting type behaviors. The term 'enhanced diffusion'
is used to describe these systems which are quite ubiquitous in nature. Phenomenon such as MHD \cite{gaw}, turbulence \cite{pommois},
and stock market indices \cite{skjeltorp} are just some of the many examples.

Here one may ask, if FBM is clearly not sufficient to describe more deterministic processes, then what paradigm 
can give a better description? The answer is; no generalized statistical picture exists which fully describes this region of enhanced diffusion.
However, based on the ubiquity of bursting and intermittent type phenomenon in nature, we can look at the scaling behavior 
of a class of distributions known as the 'stable distributions'. One of these distributions, the L\'{e}vy distribution \cite{lev}, has 
been used successfully in describing a number of enhanced diffusive processes \cite{skjeltorp,alle,scaf}. Let us digress a bit in order to examine the
scaling behavior of the L\'{e}vy distribution. 

The generalized central limit theorem provides for the generalization of the gaussian distribution into stable distributions 
of which the gaussian itself is a limiting case. The leptokurtosis or heavy tailed behavior of these stable distributions 
accounts for the bursting and sharply peaked behavior of non-gaussian phenomena in that the decay of such systems
is much slower than the gaussian. The inverse fourier transform of the characteristic function contains a number of parameters 
which describe the overall scaling invariance;
\begin{equation}
F\left(x;\alpha ,\beta  \right) = \frac{1}{Pi}\int\limits_{0}^{\infty }\exp \left(-x^{\alpha } \right)\cos \left[x t + \beta x^{\alpha }\omega \left(x,\alpha  \right)\right]dx
\label{stable}
\end{equation}
where $ \alpha $, the characteristic exponent, measures the thickness of the tails. For a set of stable random variables, larger values of $ \alpha $ imply a lower probability of observing any one of these variables since the probability density is more centrally weighted. Smaller values of $ \alpha $ dictate that the probability density is more heavily weighted  in the tails of the distribution. Note, that for the larger value of $ \alpha $ = 2, the gaussian results, making it less likely
to observe a random fluctuation which is far from the center of the distribution. The symmetry parameter, $ \beta $, is defined in the range $ -1\leq \beta \leq 1$, and governs the symmetry properties of the distribution. Closed forms of the L\'{e}vy stable distribution are known only for a limited number of $ \alpha $ and $ \beta $ parameters.  Other closed forms
are the Cauchy $ \left(\alpha = 1\right)$ and the Pearson $ \left(\alpha= \frac{1}{2}\right) $ distributions. When equation \ref{stable} is a closed form L\'{e}vy  $ \left(P_{L} = F\right)$, self-similarity is assured through the scaling relation; 

\begin{equation}
\tilde{P}_{L}\left(S_{n} \right)\equiv P_{L}\left(S_{n}\right )n^{- \alpha }. 
\label{selsim}
\end{equation}
and the asymptotic form of $P_{L}$ for large values of x is an inverse power law:
\begin{equation}
P_{L\left(\mid x\mid  \right)}\sim \mid x\mid. ^{- \left(1+ \alpha  \right)}
\end{equation}
The asymptotic behavior of this distribution results in the consequence that the moments of the distribution, $ E\left(\mid x\mid ^{n} \right)$, diverge for
$ n\geq \alpha $ when $ \alpha < 2$. Hence this distribution lacks a characteristic scale. The asymptotic form of the Levy distribution can 
evidence power law exponents strongly dictated by the intermittent or stochastic properties of the dynamical system it represents, for example,
Zumofen and Klafter \cite{Zum} evaluated the diffusive properties of intermittent systems with the result that L\'{e}vy statistics
occurs in a power law region, $  2< \mu< 3 $ , when the scaling parameter, 
\begin{equation}
\delta = \frac{1}{\left(\mu - 1 \right)}.
\label{svm}
\end{equation}

This scaling relation is derived from a coupled space-time probability based on the velocity walk and jump models. 
The inclusion of a coupled space-time memory is what accounts for enhanced diffusion effects such as chaotic intermittency.
When $ \mu = 3 $, the scaling parameter, 
$ \delta = 0.5$ is representative of a purely random noise situation in which no memory correlations exists. When $ \mu = 2 $, 
the ballistic peak limit  ($\delta = 1$) of the L\'{e}vy distribution is reached, below which ($\mu < 2$) the scaling relation no longer applies. 
In essence, one end of this region is characterized by increasing 
gaussian behavior with the other end characterized by an increasing L\'{e}vy statistical signature indicating long term memory correlations.
Grigolini, et. al. \cite{grigo}, in redefining the diffusion process as an asymmetric one found the same scaling
relation as Eq. (\ref{svm}). The effect of representing the asymmetric process with an asymmetric jump model, was to better 
represent the biased direction of a time series containing either positive or negative values for data points. 
In looking at the empirical fits values of $ \mu $ seen in figures 3 and 4, we note that both of the values exceed 2.0, falling within the 
$2.0< \mu  < 3.0$ region. Therefore, based on the knowledge that the region, $ 0.5 < H < 1.0$, reflects increasing memory correlations as $ H\to 1$, and a 
random gaussian (no memory) at $ H= 0.5$,
we make the {\it ansatz} that the persistant random walk of FBM can be replaced by the asymmetric jump model as a better representation of
our data. Therefore, we examine the statistics of the waiting times distributions by setting our scaling parameter, H equal to
$ \delta $ of equation 11 and solve for the corresponding power law exponents, $ \mu $. Then, we compare these values to the $ \mu $ values 
of our empirical fits (denoted as $ \mu _{fit}$). Tables 2  summarizes our results.
We note that the values of $\mu = 2.39 $ for $ \psi _{I}\left(t \right)$ calculated by 
using the L\'{e}vy scaling equation for $ \delta $ = H, matches the corresponding empirical fit, $\mu_{fit} = 2.39\pm 0.03 $ 
and $ \mu = 2.27$ for $ \psi _{B\left(t \right)}$ falls within the error window of $\mu_{fit} = 2.29\pm 0.02 $. This close agreement
between the $\mu$ values indicates that the the identification of the hurst scaling parameter, H, 
with the $\delta$ scaling parameter of the asymmetric jump model is a significant one in that we can identify the waiting
time statistics as other than the gaussian statistics of fractional brownian motion. This indicates that the temporal intermittency of 
flaring/bursting processes is best characterized by L\'{e}vy statistics.

\section{Summary of conclusions}
In this study, the Hurst scaling parameter was utilized under the assumptions of the FBM/CTRW
framework and the scaling relationship of Eq. (\ref{svm}) to draw some important conclusions regarding the nature of the hard X-ray
flaring process. First, the deviation of the $ \psi _{\left(t \right)}$  scaling parameter, H, from the random 
gaussian regime (H = 0.5)
provides clear evidence that the statistical properties
which represent the underlying dynamics of the waiting times 
cannot be adequately represented by the randomly distributed fluctuations of ordinary Brownian motion, i.e.;
the Gaussian and it's related distribution, the Poisson. Memory correlations inherent in the waiting time 
dynamics are of a persistant nature in that the diffusion of these systems is likely to be characterized by bursting and/or intermittent type 
behaviors propagated in the same way. Finally, the waiting time distributions fall into a region of enhanced diffusion which is 
best characterized by L\'{e}vy statistics.

\section{Acknowledgements}
The author wishes to thank Jim McTiernan and M. Wheatland for generously supplying the ICE/ISEE3 hard X-ray
data and the BATSE/CGRO NASA science team who made possible, the solar flare catalog list from which the 
BATSE data was obtained. I am most grateful to my advisor, Dr. Paolo Grigolini, for his guidance and suggestions, 
and to Markus Aschwanden for many enlightening discussions and suggestions. In addition, I would like to thank
Heikki Ruskeepaa and Atul Sharma for their generous help with Mathematica programming.

\newpage
\begin{table}[t]
\begin{center}{{\bf Table 1}: \\Hurst parameters and p values for $ \psi \left(t \right)$.  } 
 \begin{tabular}{|c|c|c|}\hline
        Flare Parameter  &  H    &   p value   \\ \hline\hline
        $ \psi _{I}\left(t \right)$ &   $ 0.72\pm 0.01$   &  3.77 x 10$^{-14}$    \\ \hline
         $ \psi _{B}\left(t \right) $ &   $ 0.79\pm 0.01$   &    2.22 x 10$^{-16}$  \\ \hline
  \end{tabular} \\*[4cm]
\end{center}
\end{table}

\begin{table}[h]
\begin{center}{{\bf Table 2}: \\ Hurst parameters and $ \psi \left(t \right)$ 
distribution exponents, $ \mu $.} \\
 \begin{tabular}{|c|c|c|c|}\hline
        Flare Parameter  &  H    &   $ \mu _{fit}$ &  $ \mu $ \\ \hline\hline
        $ \psi _{I}\left(t \right)$ &   $ 0.72\pm 0.01$   &  $2.39\pm 0.03$  & 2.39  \\ \hline
        $ \psi _{B}\left(t \right) $ &   $ 0.79\pm 0.01$   &    $2.29\pm 0.02$ & 2.27  \\ \hline
  \end{tabular} \\ 
\end{center}
\end{table}

\newpage
\begin{figure}
\begin{center}
\epsfig{file=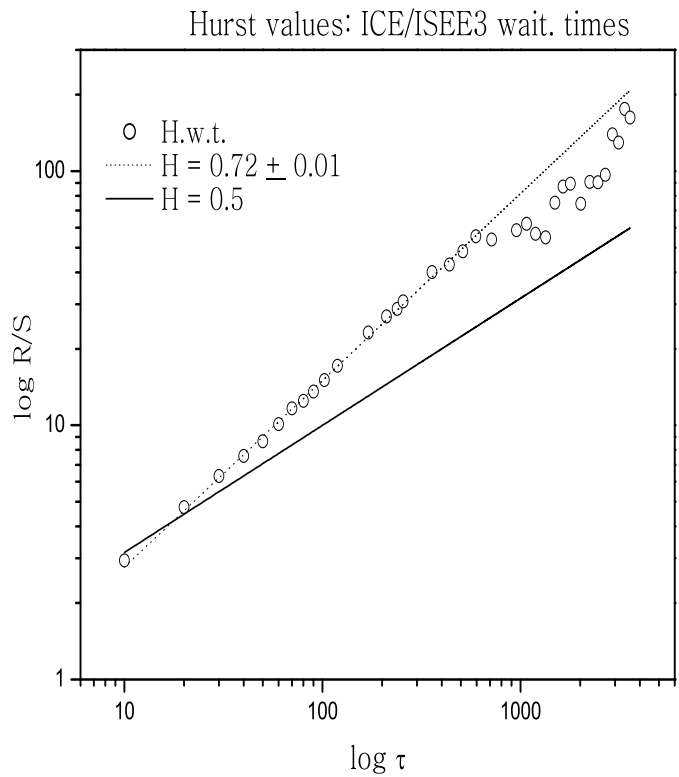,height=12cm,width=10cm,angle=0}
\end{center}
\caption{ \small{Rescaled range results: ICE/ISEE3 Waiting Times. Windows of the data were obtained through whole number 
divisors of the data length less 4 (3574 - 4 = 3570]. A least squares regression fit was performed from the 2nd through the 19th
window out of a total of 35 windows.}}
\end{figure}

\begin{figure}
\begin{center}
\epsfig{file=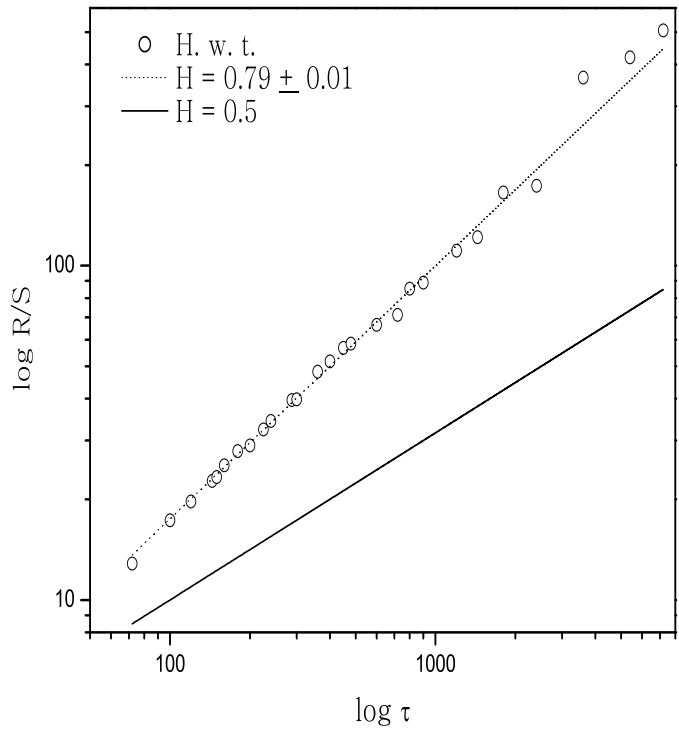,height=12cm,width=10cm,angle=0}
\end{center}
\caption{ \small{Rescaled range results: BATSE Waiting Times. A total of 27 windows of the data were obtained from
whole number divisors of the data length less 12 (7212 - 12 = 7200). A least squares linear fit was performed from
the 2nd to the 19th window leaving a leftover of 8 windows due to poor statistical fluctuations within the larger
windows. The H = 0.5 solid line representing random gaussian statistics and scaled to the minimum window value
is included for reference purposes. }}
\end{figure}

\begin{figure}
\begin{center}
\epsfig{file=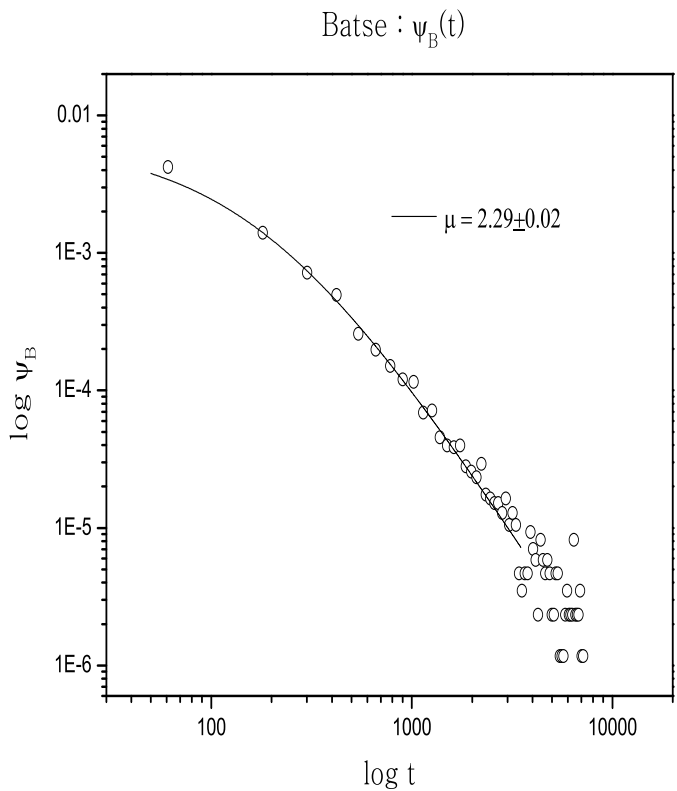,height=12cm,width=10cm,angle=0}
\end{center}
\caption{\small{Log-Log plots: BATSE Waiting times distributions, $\psi _{B}\left(t \right)$. 
The frequency distribution was fitted by using the asymptotic power law
given by Eq. (\ref{invpwr}).The regression is based on a least squares fit to the inverse power law
basis functions where the coefficients are the parameters of the fit. An elimination of the 1st
five bins and binned data close to the abscissa value of $ 10^{4}$ improved the fit routine which was then
applied to the binned data between the 2nd and the 25th bins.} }
\end{figure}

\begin{figure}
\begin{center}
\epsfig{file=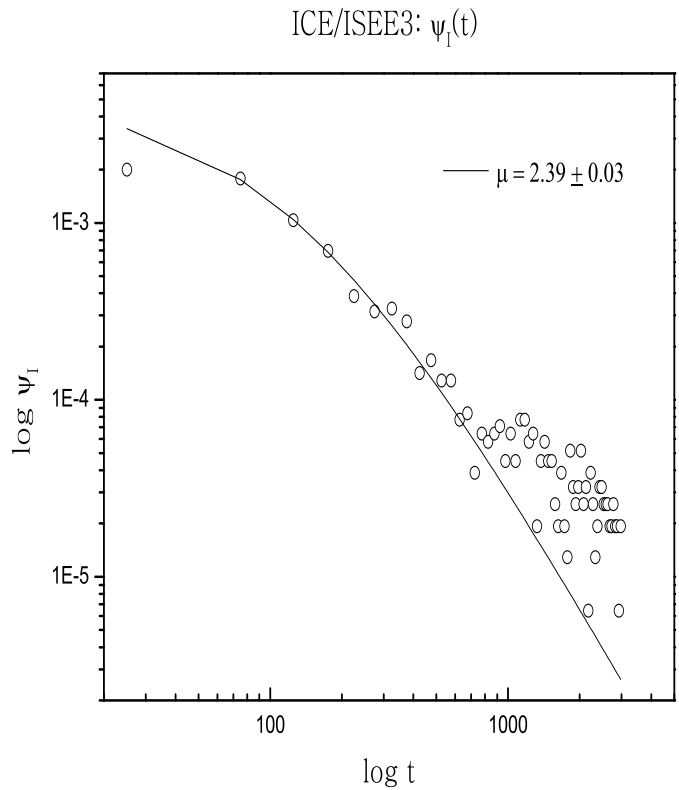,height=12cm,width=10cm,angle=0}
\end{center}
\caption{\small{Log-Log plots: ICE/ISEE3 Waiting times Distributions, $ \psi _{I}\left(t \right)$.
The asymptotic power law
given by Eq. (\ref{invpwr}) was used to fit the distribution. The regression technique employed is the same as the one used 
for the BATSE binned data and involved the first 17 binned data points.}}
\end{figure}
\end{document}